\documentclass[pdftex,epjc3]{svjour3}
\RequirePackage[T1]{fontenc}

\smartqed  % flush right qed marks, e.g. at end of proof

\RequirePackage{mathptmx}      % use Times fonts if available on your TeX system
\RequirePackage{flushend}
\RequirePackage[numbers,sort&compress]{natbib}
\RequirePackage[colorlinks,citecolor=blue,urlcolor=blue,linkcolor=blue]{hyperref}

\usepackage[utf8]{inputenc}
\usepackage{amsmath,amssymb,amsbsy,amstext,amsfonts}
\usepackage{graphicx,multicol}
\usepackage[svgnames,dvipsnames,x11names]{xcolor}
\usepackage{hyperref}

\journalname{The European Physical Journal C}

\def\beq{\begin{equation}}
\def\eeq{\end{equation}}

\def\bea{\begin{eqnarray}}
\def\eea{\end{eqnarray}}

\begin{document}

\title{$\alpha$-attractor from superconformal E-models in brane inflation }
\author{Mudassar Sabir\thanksref{e1,addr1} \and Waqas Ahmed\thanksref{e2,addr2} \and \\ Yungui Gong\thanksref{e3,addr1} \and Yizhou Lu\thanksref{e4,addr1} }

\institute{School of Physics, Huazhong University of Science and Technology, Wuhan, Hubei 430074, China \label{addr1} \and
School of Physics, Nankai University, 94 Weijin Road, Nankai District, Tianjin 300071, China \label{addr2}  }

\thankstext{e1}{msabir@hust.edu.cn}
\thankstext{e2}{waqasmit@nankai.edu.cn}
\thankstext{e3}{yggong@mail.hust.edu.cn (Corresponding author)}
\thankstext{e4}{louischou@hust.edu.cn}

\date{Accepted: 23 December 2019}

\maketitle

\abstract{In the large extra dimensional braneworld inflation, Friedmann equation is modified to include a quadratic term in energy density
with an additional parameter $\lambda$ called brane tension in addition to the usual linear term. The high energy brane corrections modify the slow-roll parameters and affect the behaviour of inflation. We analyse the superconformal inflation for E-models and find that there exist $\alpha$-attractors in brane inflation.
The predictions for the scalar spectral index $n_s$ and the tensor-to-scalar ratio $r$ are computed numerically, and approximate analytic formulas in the high energy limit have been given for the observable $n_s$ and $r$. The constraints on the model parameters are obtained by using Planck 2018 and BICEP2 observational data.}

\section{Introduction}\label{sec:Intro}

The theory of cosmic inflation was put forward in the early 1980s to solve the horizon problem of the big bang theory and explain the observed large scale homogeneity and isotropy of the universe \cite{Guth:1980zm,Starobinsky:1980te,Sato:1980yn,Linde:1981mu,Albrecht:1982wi}.
Furthermore, the quantum fluctuations of the inflaton provide the seeds for the large scale structure of the universe \cite{Guth:1982ec}.
A plethora of single field inflationary models have been constructed \cite{Martin:2013tda} and new models continue to be proposed.
For example, the recently proposed constant roll inflation \cite{Martin:2012pe,Motohashi:2014ppa,Yi:2017mxs,
Gao:2018cpp,Gao:2018tdb,Gao:2019sbz} may be used to
generate large density perturbation at small scales to seed the formation of primordial black holes \cite{Germani:2017bcs,Motohashi:2017kbs,Kannike:2017bxn,Gong:2017qlj,Lu:2019sti}.
The constraints $n_s=0.965\pm 0.004$ (68\% C.L.)
and $r_{0.05}<0.06$ (95\% C.L.) \cite{Akrami:2018odb,Ade:2018gkx}
given by the temperature and polarization
measurements on the cosmic microwave background anisotropy can be used to
filter inflationary models. The observed central value of
$n_s$ suggests the relation $n_s=1-2/N$ with $N=60$, where
$N$ is the number of $e$-folds before the end of inflation.

The Starobinsky model $R+R^2$
and the Higgs inflation with the nonminimal
coupling $\xi\psi^2 R$ in the strong coupling limit $\xi\gg 1$
give the same scalar spectral tilt $n_s=1-2/N$ and
the tensor-to-scalar ratio $r=12 /N^2$
\cite{Starobinsky:1980te,Kaiser:1994vs,Bezrukov:2007ep}.
The general T-models and E-models give the
$\alpha$-attractors $n_s=1-2/N$ and $r=12\alpha /N^2$ \cite{Kallosh:2013yoa,Kallosh:2013hoa,Kallosh:2013maa}.
Furthermore, inflationary attractors can be obtained with more
general potentials and nonminimal coupling
in the strong coupling limit \cite{Kallosh:2013tua,Galante:2014ifa,Yi:2016jqr,Gao:2017uja}.
These attractors are consistent with current observations
and can be tested in the future by LiteBIRD \cite{Suzuki:2018cuy}.

Inflation is believed to have taken place at the GUT scale
or below \cite{Ahmed:2018jlv} and is assumed to be
described by the low-energy effective field theory \cite{Cheung:2007st}. To be compatible with quantum gravity,
sub-Planckian field excursion $\Delta\phi<1$ in reduced Planck unit is favoured
as the distance conjecture suggested \cite{ArkaniHamed:2006dz,Ooguri:2006in}.
To explain the observed flat universe, for single field slow-roll inflation
we require $N \gtrsim 50$ and
the needed field excursions in reduced Planck units are related to $r$ by
the Lyth bound \cite{Lyth:1996im} as $\Delta\phi \simeq N \sqrt{r/8}$ \footnote{A modified bound was discussed in \cite{Gao:2014pca}.}. However, the usual 4-dimensional single field chaotic inflationary models generally need super-Planckian field excursion $\Delta\phi>1$
in reduced Planck unit that can lead to the breakdown of effective field description since then the quantum gravity corrections can not be ignored.

In braneworld cosmology, our 4-dimensional world
is embedded in a higher 5-dimensional bulk.
Employing this setup of 5-dimensional Randall-Sundrum model II \cite{Randall:1999vf},
brane inflation was introduced \cite{Maartens:1999hf,Cline:2007yq}.
In brane inflation the Friedmann equation gets a $\rho^2$ correction
and the slow-roll parameters get modified with an extra parameter
of brane tension $\lambda$ that relates four-dimensional Planck scale $M_4$ and five-dimensional Planck scale $M_5$ as $\lambda \sim M_5^6/M_4^2$.
The modified Friedmann equation helps solve the problem of super-Planckian field excursions because the Lyth bound does not hold \cite{Lin:2018kjm,Brahma:2018hrd}.
Since the Hubble parameter $H$ in brane scenario is
larger and the Hubble damping is more effective,
brane inflation can be worked out with even steeper potentials \cite{Maartens:1999hf,Cline:1999ts,Csaki:1999jh,Ida:1999ui,Gong:2000en,Copeland:2000hn}.
For the chaotic inflation with power law potential $V=V_0 \phi^p$,
sub-Planckian field excursion can be easily satisfied for small $p$ \cite{Lin:2018kjm}.
Besides, the brane scenario has been shown to naturally solve the hierarchy problem of weak scale as well as weakness of gravity.
This makes the brane inflation interesting to explore \cite{Jaman:2018ucm,Safsafi:2018cua,Es-sobbahi:2018yfh,Bhattacharya:2019ryo,Jawad:2019hzo,Sabir:2019xwk,Sabir:2019bsh}.

The inflationary attractors and brane inflation are both interesting, so it is worthy
of discussing the existence of attractors in brane inflation.
In this paper, we use the superconformal E-models to study attractors in brane inflation. The Planck 2018 and BICEP2 data \cite{Akrami:2018odb,Ade:2018gkx} are used to constrain the model parameters. The sub-Planckian field excursion is also
discussed. The paper is organized as follows. In Sec. \ref{sec:brane}, we briefly review the brane inflation.
In Sec. \ref{sec:Attractor}, the $\alpha$-attractors for the superconformal E-models are obtained both numerically and analytically in the high energy limit.
The conclusions are presented in Sec. \ref{sec:Conclusion}.

\section{Brane inflation}\label{sec:brane}

In the braneworld cosmology our four-dimensional world is a 3-brane embedded in a higher-dimensional bulk. The Friedmann equation in braneworld picture gets a $\rho^2$ correction term as \cite{Csaki:1999jh,Maartens:1999hf,Binetruy:1999hy,Binetruy:1999ut}
\begin{equation}
H^2=\frac{1}{3M_P^2}\left(\rho+\frac{\rho^2}{2 \lambda} \right),\label{eq:Friedmann}
\end{equation}
where $\lambda$ is the brane tension that relates four-dimensional Planck scale $M_4$ and five-dimensional Planck scale $M_5$ as
\begin{equation}
\lambda=\frac{3}{4\pi} \frac{M_5^6}{M_4^2}~,
\end{equation}
where $M_P=M_4/\sqrt{8\pi}=1$ is the reduced Planck mass. The nucleosynthesis limit implies that $\lambda \gtrsim (1 \mbox{ MeV})^4 \sim (10^{-21})^4$ in reduced Planck units. A more stringent constraint can be obtained by requiring the theory to reduce to Newtonian gravity on scales larger than 1 mm corresponding to $\lambda \gtrsim 5 \times 10^{-53}$ i.e. $M_5 \gtrsim 10^5$ TeV \cite{Maartens:1999hf}. Notice that in the limit $\lambda \rightarrow\infty$ we recover standard Friedman equation in four dimensions.

The $\rho^2$ term in the modified Friedmann equation changes the definitions of the slow-roll parameters as \cite{Langlois:2000ns}
\begin{eqnarray}
\epsilon &=& \frac{1}{2} \left( \frac{V^\prime}{V} \right)^2 \frac{1}{\left( 1+\frac{V}{2\lambda} \right)^2}\left( 1+\frac{V}{\lambda} \right), \label{eq:epsilon} \\
\eta &=& \frac{V^{\prime\prime}}{V} \left( \frac{1}{1+\frac{V}{2 \lambda}} \right). \label{eq:eta}
\end{eqnarray}
Accordingly the formula for the number of $e$-folds before the end of inflation becomes
\begin{equation}
N(\phi) = -  \int_{\phi}^{\phi_{\rm end}}\frac{ V }{ V'} \left(
1+{ V\over 2\,\lambda}\right) \mathrm{d}{\phi}~.\label{eq:efolds}
\end{equation}
In the low energy limit $V/\lambda\ll 1$ or $\lambda\rightarrow \infty$,
the usual definitions in standard cosmology are recovered.
However, in the high energy limit $V/\lambda\gg 1$, the slow-roll parameters
$\epsilon$ and $\eta$ are suppressed by a factor $V/\lambda$, so even a steep
potential easily satisfies the slow-roll conditions.

For the Randall-Sundrum model II \cite{Randall:1999vf},
the amplitudes for the tensor and scalar power spectrum are \cite{Maartens:1999hf,Langlois:2000ns}
\begin{eqnarray}
A_t^2 &=&\frac{2}{3 \,\pi^2}\,  V \, \left( 1+\frac {V}{ 2\, \lambda}\right)F^2~, \label{eq:Pt}\\
A_s^2 &=& \frac{1}{12 \,\pi^2}\,\frac{V^3}{V^{\prime2}} \left( 1 + \frac{V}{2\,\lambda} \right)^3 ~, \label{eq:Ps}
\end{eqnarray}
where
\begin{equation}
F^2=\left[\sqrt{1+x^2} -x^2 \sinh^{-1}\left({1\over x}\right)\right]^{-1}~,\label{eq:F2}
\end{equation}
and
\begin{align}
\label{eq:x}
x\equiv \left(\frac{3 \,H^2}{4 \pi\,\lambda}\right)^{1/2} = \left[\frac{2\,V} {\lambda}\left(1+\frac{V}{2\,\lambda}\right)\right]^{1/2}~.
\end{align}
In the low-energy limit $V/\lambda\ll 1$, $F^2\approx 1$, we recover
the results in standard cosmology \footnote{The results \eqref{eq:Pt} and \eqref{eq:Ps} differ from those in Refs. \cite{Maartens:1999hf,Langlois:2000ns}
by factors of $(1/16)\times (4/25)$ and $4/25$ respectively
because they use matter density perturbations when modes re-enter the Hubble scale during
the matter dominated era.}.

The spectral tilt for scalar perturbation can be written in terms of the
slow-roll parameters as \cite{Maartens:1999hf}
\begin{equation}
\label{eq:ns}
n_{s} - 1  = -6\epsilon + 2\eta,
\end{equation}
and the tensor-to-scalar ratio is \cite{Bento:2008yx}
\begin{equation}
\label{eq:r}
r = \frac{A_t^2}{A_s^2}=8\left(\frac{V'}{V}\right)^2\frac{F^2}{[1+V/(2\lambda)]^2}.
\end{equation}
In the low-energy limit $V/\lambda\ll 1$, $F^2\approx 1$, we recover the standard result $r=16\epsilon$. In the high-energy limit $V/\lambda\gg 1$, $F^2\approx 3\,V/2\,\lambda$,
the tensor-to-scalar ratio is modified as $r=24\epsilon$ and the amplitude for the scalar power spectrum becomes
\bea
A_s^2 &=& \frac{1}{12 \,\pi^2}\,\frac{V}{\epsilon} \left( \frac{V}{2\,\lambda} \right)^2~.
\eea
Hence in general the energy scale of inflation and the brane tension
cannot be fixed by the observational values of $A_s^2$ and $r$.

\section{Superconformal E-models}\label{sec:Attractor}

In Ref. \cite{Kallosh:2013hoa}, a broad class of chaotic inflationary models with spontaneously broken superconformal invariance was generalized to the class of cosmological attractors with a single inflaton field.
In Ref. \cite{Kallosh:2013yoa}, the phenomenology of $\alpha$-attractor from superconformal E-models was discussed.
The potentials for general E-models have the following generic form \cite{Carrasco:2015rva},
\bea
\label{eq:attractor}
V(\phi)&=&V_0\left[1-\exp \left(-\sqrt{\frac{2}{3 \alpha }}\phi\right)\right]^{2n}\,, \quad n\,=\,1,2,3,\ldots .
\eea
For small field $\phi\ll 1$, the potential becomes power law potential $V\sim \phi^{2n}$. For large field $\phi\gg 1$,
the predictions from these potentials converge to the universal attractors $n_s = 1-2/N$
and $r = 12\alpha /N^2$ in the large $N$ limit irrespective of the value of $n$.
The sub-Planckian field excursion requires that $\alpha<\sqrt{2/3}$.
However, in brane inflation the predictions change drastically as we will see below.
Since we are interested in attractors, we consider large fields only.

For a particular E-model with fixed $n$ we have three parameters: $\alpha$, $V_0$, and $\lambda$.
On the other hand, the slow-roll parameters in Eqs. (\ref{eq:epsilon}) and (\ref{eq:eta}) are functions of $\alpha$ and $V_0/\lambda$ only,
so the scalar spectral tilt $n_s$ and the tensor-to-scalar ratio $r$ also depend only on $\alpha$ and $V_0/\lambda$.
To take an example, we discuss the case with $n=1$.
For convenience, we introduce
\begin{equation}
  w(\phi)=\frac{V(\phi)}{\lambda}.
\end{equation}
As discussed in the previous section, in the limit of large brane tension $\lambda\rightarrow \infty$ or in the low energy limit $V/\lambda\ll 1$,
the effect of brane correction is negligible and the standard cosmology is recovered.
In particular, taking the large $N$ limit we get $n_s=1-2/N$ and $r=12\alpha/N^2$.
Therefore, in order to discuss brane inflation,
we consider the high energy limit $V/\lambda\gg 1$ in this section.
Then we express $\epsilon$ and $\eta$ in terms of $w(\phi)$
with $w(\phi)\gg 1$ as
\begin{equation}
\begin{split}
  \epsilon & \approx \left(\frac{w'}{w}\right)^2\frac{2}{w},\\
  \eta &\approx \frac{w''}{w}\frac{2}{w}.
\end{split}
\label{brepsilon}
\end{equation}
The $e$-folding number in the high energy limit is given by
\begin{equation}
  {\rm d}N(\phi)=\frac{w^2}{2w'}{\rm d}\phi.\label{eq:Nefolds}
\end{equation}
The amplitudes for the scalar and tensor power spectra become
\begin{align}
  A_s^2&=\frac{Vw^3}{96\pi^2}\left(\frac{w}{w'}\right)^2,\\
  A_t^2&=\frac{Vw^2}{2\pi^2},
\end{align}
and the tensor-to-scalar ratio $ r=16A_t^2/A_s^2=24\epsilon $.
Since in brane inflation the slow-roll
parameter $\epsilon$ in equation \eqref{brepsilon}
has an additional factor $1/w$, models with steeper potentials that have
$V'/V>1$ can satisfy the observational constraint $r<0.06$.

From the acceleration equation
\bea
\label{ddota}
\frac{\ddot{a}}{a}=\frac{1}{3M_P^2}
\left[(V-\dot{\phi}^2)+\frac{\dot{\phi}^2+2V}{8\lambda}
(2V-5\dot{\phi}^2)\right],
\eea
we find that inflation ends around $\dot{\phi}_{\rm e}^2\simeq 2V/5$ in the high energy regime.
This condition gives $\epsilon(\phi_{\rm e})=6/5$ which is more accurate than the usual condition $\epsilon(\phi)=1$.
Therefore, we get
\begin{equation}
  \phi_{\rm e}=\frac{1}{\beta }\ln \left(1+2\sqrt{\frac{5}{3 w_0}} \beta\right),
\end{equation}
where $\beta=\sqrt{2/3\alpha}$ and $w_0=V_0/\lambda$.
Integrating Eq. \eqref{eq:Nefolds} we obtain
\begin{equation}
\label{neq11}
  N=\frac{w_0}{8\beta^2}\left.(e^{-2\beta\phi}-6e^{-\beta\phi}+2e^{\beta\phi}-6\beta\phi)\right|_{\phi_{\rm e}}^{\phi_*},
\end{equation}
where the subscript $*$ denotes the value at the horizon crossing.
Since $\exp(-\beta\phi_*)\ll 1$, we neglect $\exp(-2\beta\phi_*)$ and $\exp(-\beta\phi_*)$ terms in Eq. \eqref{neq11}, then we get
\begin{equation}
\label{eq:W-solution}
 e^{\beta\phi_*}\approx -3
    W_{-1}\left[-\frac13\exp\left(-\frac{4\beta^2(N+C)}{3w_0}\right)\right],
\end{equation}
where $W_{-1}(x)$ is the lower branch of the Lambert function
that satisfies the equation $W_{-1}(x)e^{W_{-1}(x)}=x$ and
\begin{equation}
  C=\frac{w_0}{8\beta^2}(e^{-2\beta\phi_{\text e}}-6e^{-\beta\phi_{\text e}}+2e^{\beta\phi_{\text e}}-6\beta\phi_{\text e}).
\end{equation}
Substituting the results into $n_s$ and $r$, we get
\begin{align}
  r(N)&=\frac{1152\lambda}{\alpha V_0}\frac{W_{-1}^2}{(3W_{-1}+1)^4},\label{eq:rN}\\
  n_s(N)&=1- \frac{48\lambda}{\alpha V_0}\frac{4W_{-1}^2-3W_{-1}^3}{(3W_{-1}+1)^4}.
  \label{eq:nsN}
\end{align}
In the large $N$ limit, $W_{-1}\approx -8 N \lambda/(9 \alpha V_0)$, we obtain
\begin{gather}
\label{nsreq11}
  n_s(N)=1-\frac{2}{N},\\
\label{nsreq12}
  r(N)=\frac{18\alpha V_0}{ N^2 \lambda}.
\end{gather}
In the high-energy limit $V_0/\lambda\gg 1$, the brane
correction increases $r$ by a factor $V_0/\lambda$,
so $V_0/\lambda$ cannot be too large. On the other hand, from Eq. \eqref{eq:W-solution} we see that the approximation
$\exp(-\beta\phi_*)\ll 1$ breaks down if $V_0/\lambda$ goes to infinity,
so the analytical approximations \eqref{nsreq11} and \eqref{nsreq12} cannot be trusted if $V_0/\lambda$ is very large.
From Eq. \eqref{nsreq12}, we see that the parameters $\alpha$
and $V_0/\lambda$ are degenerated.
For different values of $\alpha$, it is always possible to adjust
the value of $V_0/\lambda$ so that we get the same observables $n_s$ and $r$.
In other words, the observables $n_s$ and $r$ depend on the combined parameter
$\alpha'=2\alpha V_0/(3\lambda)$ only
even though there are three parameters in the model,
so the attractors in the braneworld are the same as those in standard cosmology
if we identify $\alpha'$ as $\alpha$.
Although we have two parameters $\alpha$ and $V_0/\lambda$ to vary that would have swept an area in the $n_s-r$ plane,
only the product $\alpha V_0/\lambda$ is relevant in the
high energy limit, so we expect a single curve instead of a region
in the $n_s-r$ plane when we vary both $\alpha$ and $V_0/\lambda$.
\begin{figure*}[ht]
\centering
\includegraphics[width=0.45\textwidth]{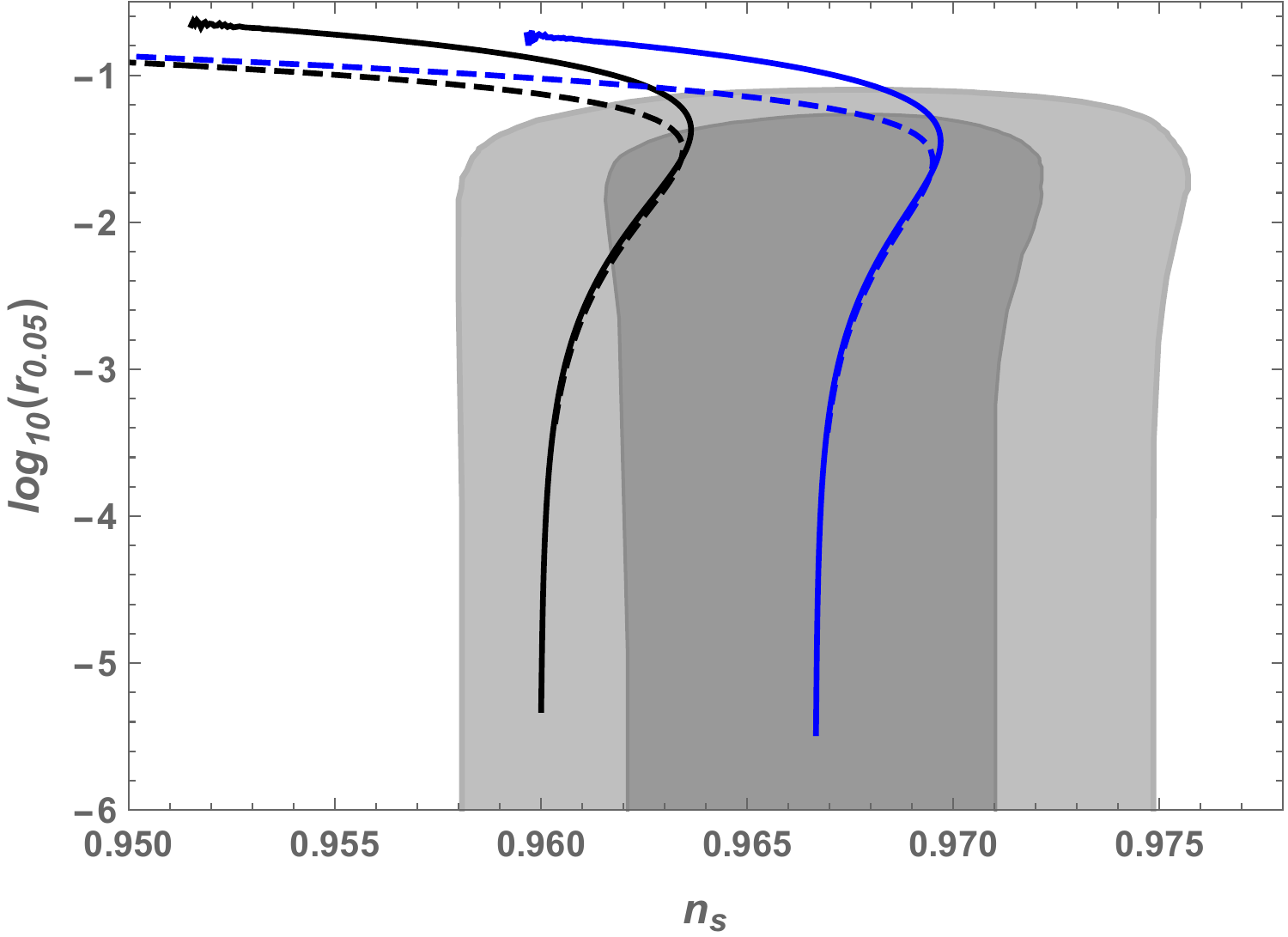}\qquad
\includegraphics[width=0.45\textwidth]{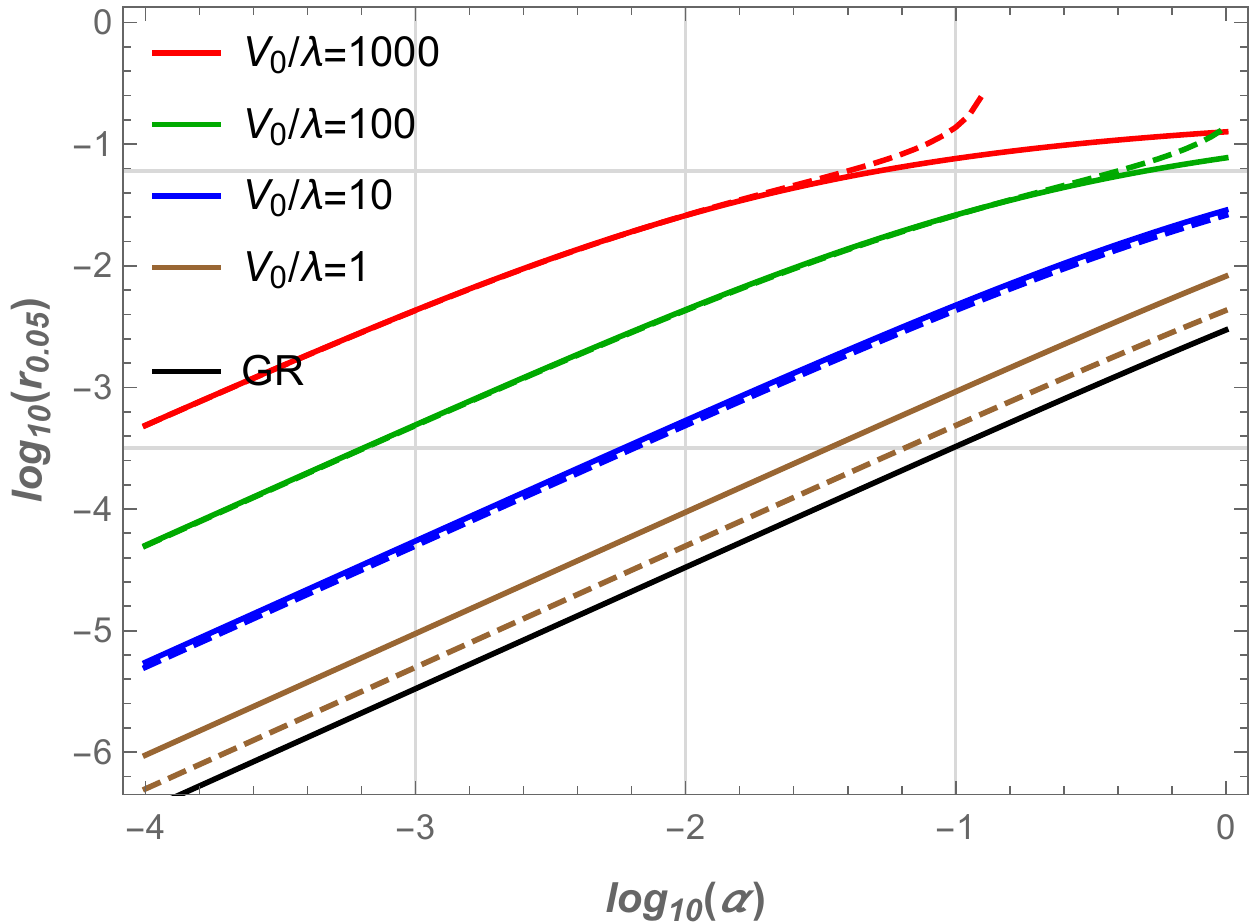}
\caption{
The left panel shows the marginalized $1\sigma$ and $2\sigma$
confidence level contours for
$n_s$ and $r$ from Planck 2018 and BICEP2 results \cite{Akrami:2018odb,Ade:2018gkx} and the results for the E-model with $n=1$ and $\alpha =0.001$ in brane inflation.
The dashed line denotes the analytical approximation \eqref{eq:rN} and \eqref{eq:nsN}
and the solid line corresponds to the numerical
result by varying the parameter $V_0/\lambda$.
The left black lines are for $N=50$ and the right blue lines are for $N=60$.
The right panel shows the relation between $r$ and $\alpha$ for different values of $V_0/\lambda$ in brane inflation and for $\alpha$-attractor in general relativity (GR) with $N=60$. The dashed line denotes the analytical approximation
and the solid line corresponds to the numerical result. The upper light gray
horizontal line represents the observational upper bound $r\leq0.06$
and the lower light gray horizontal line corresponds to the GR result with
$\alpha=0.1$.}
\label{fig:N_50_60}
\end{figure*}

To check the accuracy of the analytical approximations \eqref{eq:rN} and \eqref{eq:nsN}, we compare them with numerical results in Fig. \ref{fig:N_50_60}.
In the left panel in Fig. \ref{fig:N_50_60},
by varying $V_0/\lambda$,
we plot the analytical results with dashed lines
and the numerical results with solid lines for $N=50$ and $N=60$.
We also show the relation between $r$ and $\alpha$ for different values of $V_0/\lambda$ in the right panel in Fig. \ref{fig:N_50_60}.
From Fig. \ref{fig:N_50_60}, we see that the analytical approximations deviate from the numerical ones when $V_0/\lambda$ becomes either very large or very small.
This is easy to understand for $V_0/\lambda\lesssim 1$ as this is the low energy limit where high energy approximation is not applied.
As discussed above, for $V_0/\lambda\gtrsim 10^3$ the approximation \eqref{eq:W-solution} breaks down, so the analytical approximations are not good when $V_0/\lambda$ is either very large or very small.

One interesting point is that even if the analytical approximations are not good when $V_0/\lambda$ becomes small, it still stays on the line given by the numerical result.
The reason is due to the degeneracy between $\alpha$ and $V_0/\lambda$ in Eq. \eqref{nsreq12}, although the point $\alpha=0.01$ and $V_0/\lambda=1$
does not approximate the model well, but it actually approximates the model with $\alpha=0.001$ and $V_0/\lambda=10$ well. In the right panel of Fig. \ref{fig:N_50_60} we compare the attractors in braneworld against the standard general realtivity result for $n=1$ E-model. From the lower light gray horizontal line in the right panel,
we see that the model with different values of $V_0/\lambda$ and $\alpha$
have the same result as $\alpha$
attractors in standard cosmology with $\alpha=0.1$.
We conclude that the analytic results (\ref{eq:rN}) and  (\ref{eq:nsN}) work well in the regime
where the brane effects are appreciable i.e. $10^1\lesssim V_0/\lambda\lesssim 10^3$.
Furthermore, in both the low energy and the large $N$ limit the observables $n_s$ and $r$ become independent of $V_0/\lambda$.
Since the analytical approximations work only in a small range of $V_0/\lambda$ and only a small parts of the results for $N=50$
are consistent with the observations at the $1\sigma$ confidence level,
we use the numerical results and take $N=60$ in the following discussion.

\begin{figure}[ht]
  \centering
  \includegraphics[width=0.5\textwidth]{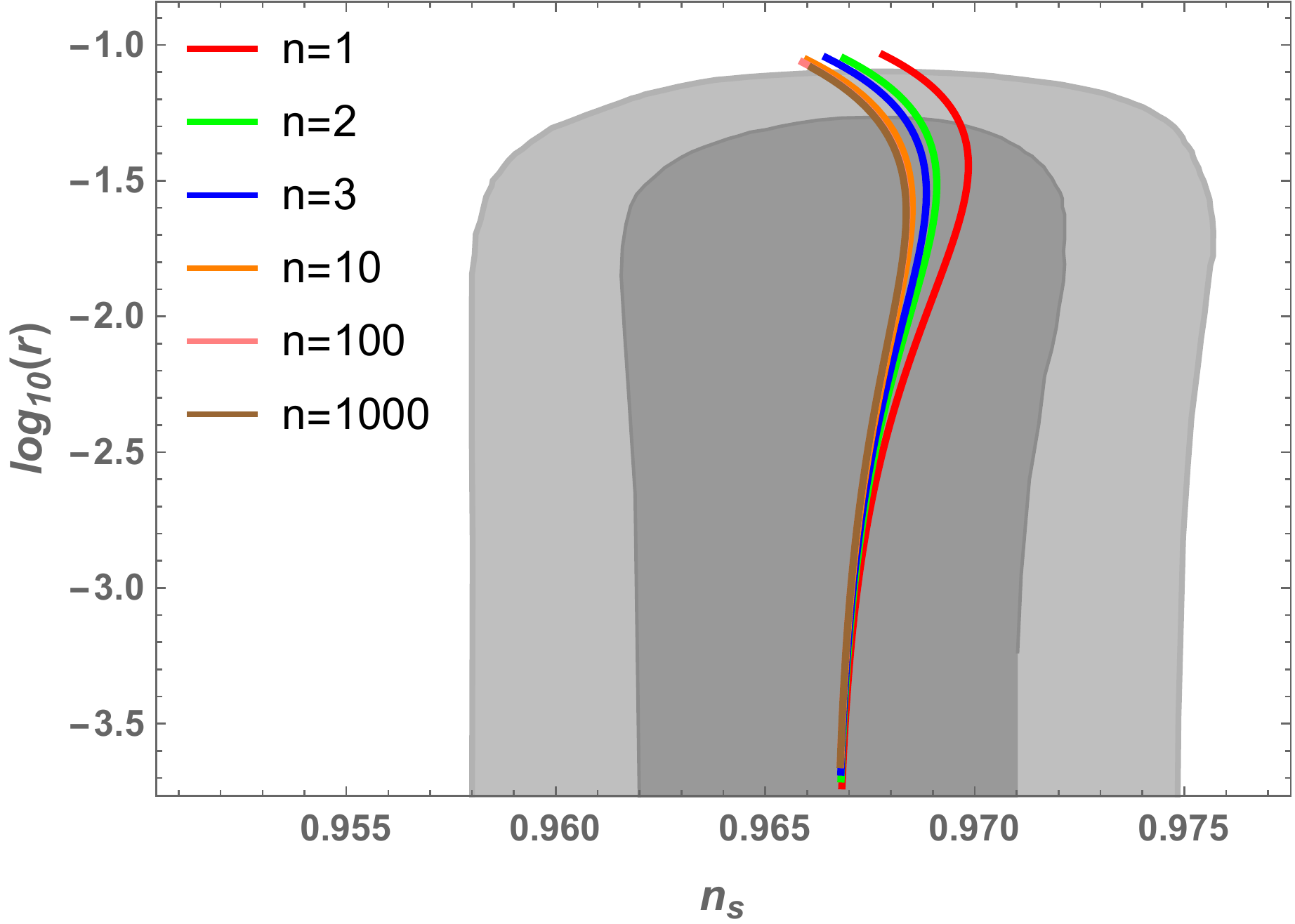}
  \caption{The marginalized $1\sigma$ and $2\sigma$ confidence level contours for
$n_s$ and $r$ from Planck 2018 and BICEP2 results \cite{Akrami:2018odb,Ade:2018gkx} and the numerical results for the $\alpha$-attractor from E-models with $n\,=\,1,2,3,10,100,1000$ in brane inflation. $\alpha$ is fixed to be $0.02$ for each model.}
  \label{fig:nsr}
\end{figure}

By choosing $\alpha=0.02$,
Fig. \ref{fig:nsr} shows the observables $n_s$ and $r$ for the E-models with $n\,=\,1,2,3,10,100,1000$.
The general behaviour is similar to $n=1$ case as we get a single curve for each model and they all converge to the attractor solution in the low energy limit.

\begin{figure*}[ht]
\centering
\includegraphics[width=0.45\textwidth]{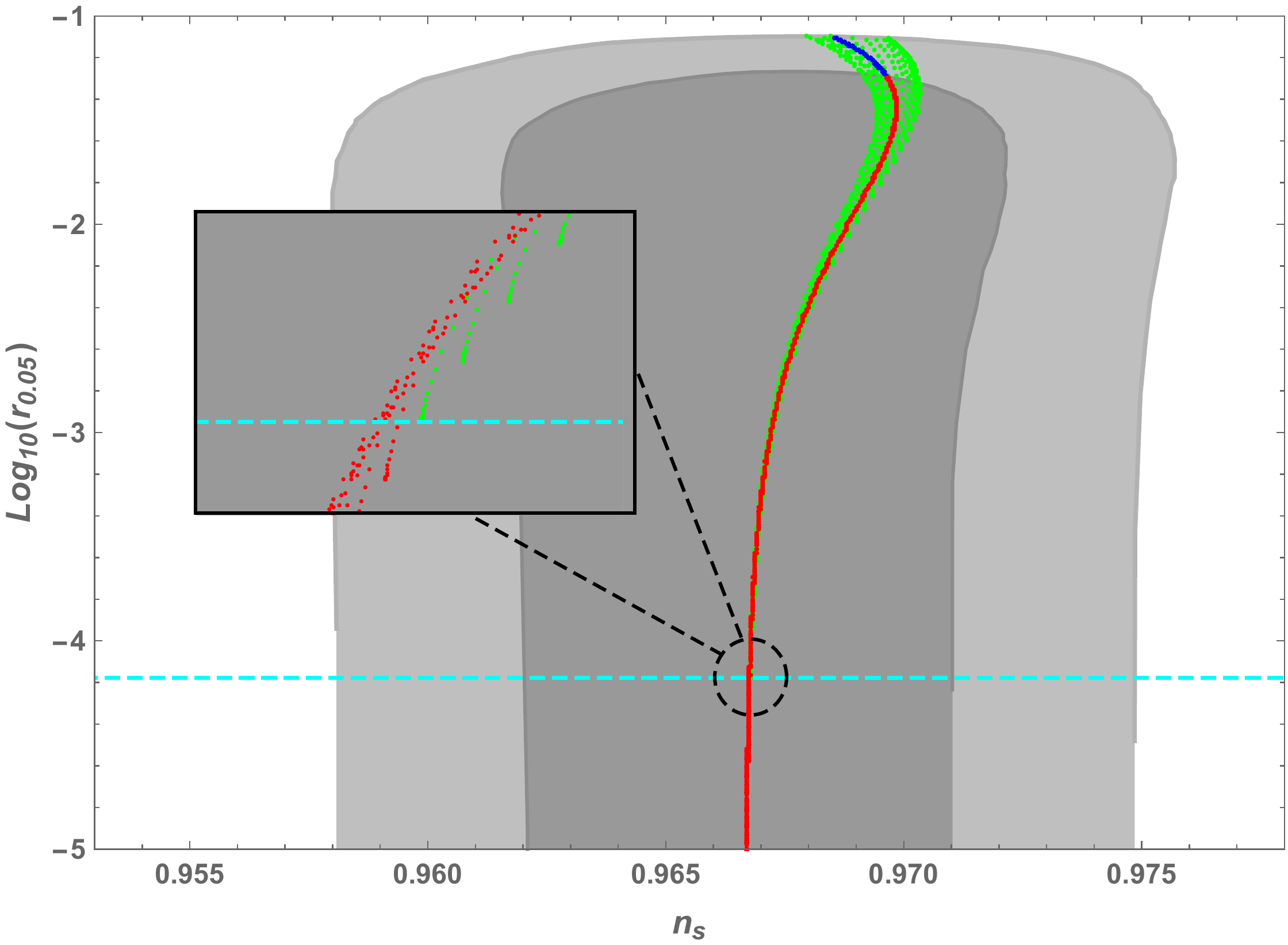}\qquad
\includegraphics[width=0.47\textwidth]{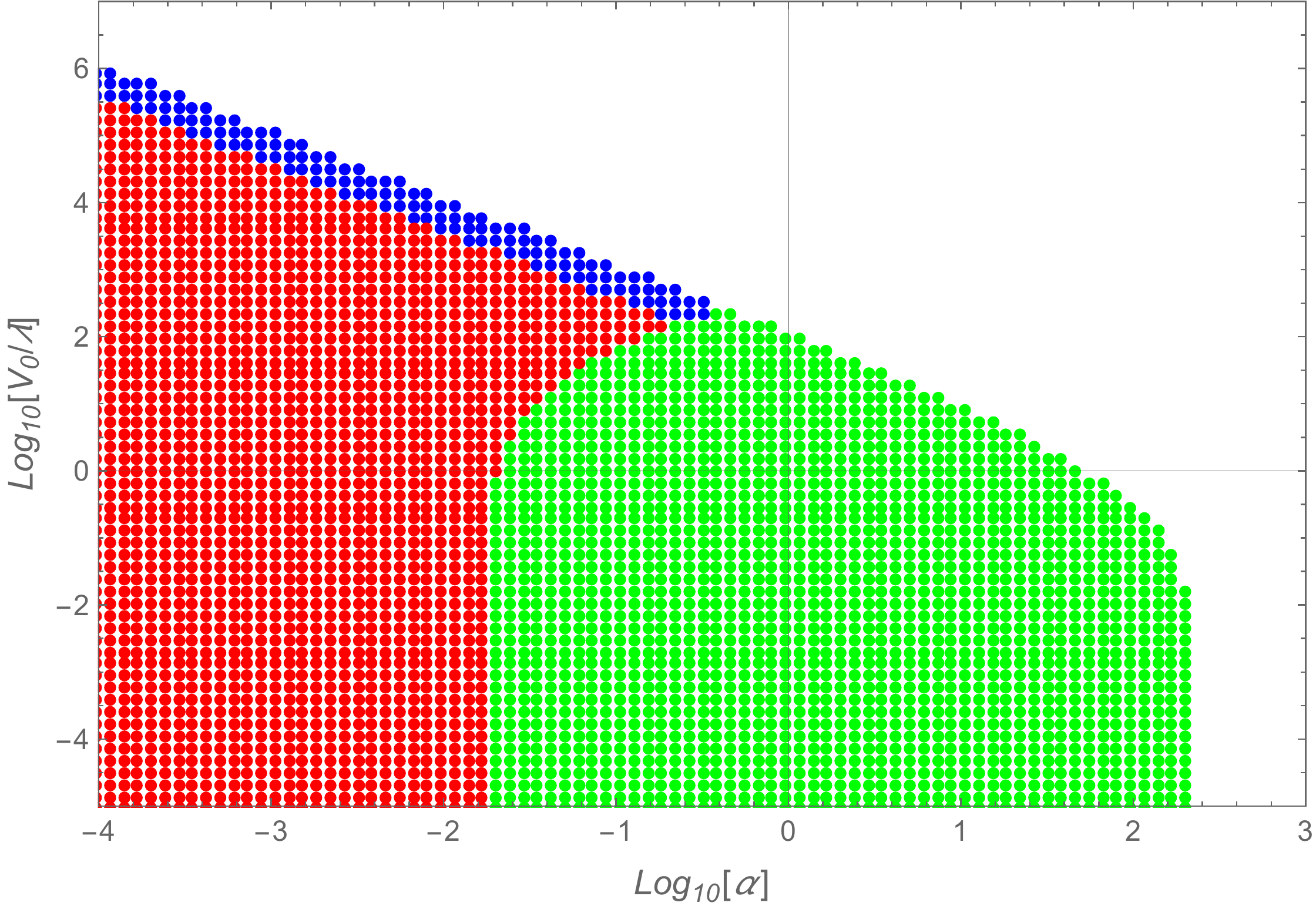}
\caption{The marginalized $1\sigma$ and $2\sigma$ confidence level contours for
$n_s$ and $r$ from Planck 2018 and BICEP2 results \cite{Akrami:2018odb,Ade:2018gkx} and the numerical results for the E-models with $n=1$ in brane inflation.
In the left panel, the cyan dashed line lies at about $r\approx 10^{-4.17}$ which
is the sufficient but not necessary upper limit for $\Delta\phi<1$,
and the inset shows the small regions circled in the figure.
The right panel shows the constraints on the parameters $\alpha$ and $V_0/\lambda$,
the red points are within $1\sigma$ region while blue points are within $2\sigma$ but outside the $1\sigma$ bounds. The green points are excluded
due to super-Planckian field excursion $\Delta\phi>1$.}
\label{fig:3}
\end{figure*}

By varying the model parameters $\alpha$ and $V_0/\lambda$, we get
the observables $n_s$ and $r$ as shown in Fig. \ref{fig:3}.
As explained above, all points are aligned almost
along a single curve, except for very large $V_0/\lambda$ where the approximation
breaks down. From Fig. \ref{fig:3}, we see that for larger value of $V_0/\lambda$,
even bigger value of $\alpha$ is allowed.
It is known that the super-Planckian field excursion can lead to the
violation of effective field description because then the quantum gravity correction
can not be ignored. Fig. \ref{fig:3} also shows the parametric space that is
constrained due to the condition $\Delta\phi<1$ for $n=1$ E-model. The unconstrained
and the constrained parametric space is colored differently. The green-colored points
correspond to $\Delta\phi>1$ while the red and blue regions lie within $1\sigma$ and
$2\sigma$ range in the $n_s$-$r$ plane and satisfy the condition $\Delta\phi<1$.
For the E-model with $n=1$, the condition $\Delta\phi<1$ implies $\alpha \lesssim 0.2$
as shown in the right panel of Fig.~\ref{fig:3}.

\section{Conclusions}\label{sec:Conclusion}

We have found inflationary attractors for the superconformal E-models
in braneworld scenario and  derived the analytical expressions for the
observables $n_s$ and $r$ in the high energy regime $V_0/\lambda\gg 1$.
The analytical approximation works well in the regime $10\lesssim V_0/\lambda\lesssim 10^3$ where the brane correction is important.
In the low energy limit $V_0/\lambda\ll 1$, the brane correction is negligible, we recover the $\alpha$-attractors $n_s=1-2/N$ and $r=12\alpha/N^2$ in standard cosmology.
In the large $N$ limit, the approximate analytic result for brane inflation in the high energy regime is the same as that in the low energy
limit if we identify the parameter $\alpha'=2\alpha V_0/(3\lambda)$ in
the high energy regime as $\alpha$ in the low energy regime,
so the brane correction requires smaller $\alpha$ to get the same $r$.
The numerical calculation of $n_s$ and $r$ shows that even
bigger values of $\alpha$ satisfy the observational
constraints if $V_0/\lambda$ becomes larger.
In standard cosmology, small $r$ means flat potential,
so large field is required for $\alpha$ attractors.
In brane inflation for E-models,
due to the factor $V_0/\lambda$ in the slow-roll parameter,
even steep parts of the potential with $V'/V>1$ give small $r$, so brane E-model
in the sub-Planckian regions ($\phi<1$) satisfies the observational constraints.
By using the observational data, we get constraints on the model parameters $\alpha$ and $V_0/\lambda$.
For the E-model with $n=1$, the sub-Planckian
field excursion condition $\Delta\phi<1$ implies $\alpha \lesssim 0.2$.

In conclusion, the E-models in braneworld scenario have $\alpha$-attractors and they satisfy the Planck 2018 and BICEP2 constraints on $n_s$ and $r$. In brane inflation,
it is easy to get sub-Planckian field excursion $\Delta\phi<1$.

\begin{acknowledgements}
This research was supported in part by the National Natural Science
Foundation of China under Grant No. 11875136,
the Major Program of the National Natural Science Foundation of China under Grant No. 11690021.
We are thankful to Zhu Yi for helpful correspondence.
MS was supported by HEC Pakistan Fellowship.
\end{acknowledgements}

\bibliographystyle{spphys}
%\bibliography{References}

\end{document}